\documentclass[a4paper,11pt]{article}
\pdfoutput=1 

\usepackage{jcappub} 

\usepackage[T1]{fontenc} 

\usepackage{natbib}
\title{Massive Population III Galaxies Near a GN-z11 Analog: The Role of the Baryon-Dark Matter Streaming Velocity}

\author[a]{Eli Visbal,}
\author[b]{Greg L. Bryan,}
\author[a]{and Ryan Hazlett}

\affiliation[a]{Department of Physics and Astronomy and Ritter Astrophysical Research Center, University of Toledo, 2801 W. Bancroft Street, Toledo, OH, 43606, USA}

\affiliation[b]{Department of Astronomy, Columbia University, 550 West 120th Street, New York, NY, 10027, USA}

\emailAdd{Elijah.Visbal@utoledo.edu}

\abstract{Recently the \emph{James Webb Space Telescope} (\emph{JWST})
detected Hebe, a Population III (Pop~III) candidate located ${\sim}3$ kpc in
projection from the massive $z=10.6$ galaxy GN-z11. This raises the question
of whether Pop~III star formation is expected in the vicinity of massive
high-redshift galaxies or whether it would be prevented by early metal
enrichment. We address this with a semi-analytic model of Pop~III and
metal-enriched star formation applied to a zoom-in N-body simulation of a
highly overdense region surrounding a GN-z11 analog. The zoom-in region resolves the
minihalos where the first stars form. We find that Pop~III and
metal-enriched star formation both occur earlier, and with a higher star
formation rate density, than in a typical mean-density environment. The
majority of Pop~III star formation in minihalos ends by $z\sim15$; however, we identify a distinct population of late-forming
Pop~III galaxies that form near the central galaxy. These systems have their star formation delayed due to ionizing feedback from the
central GN-z11 analog and experience a Lyman-Werner (LW) intensity of
${\sim}10$ -- $100~J_{21}$, potentially leading to massive Pop~III starbursts
similar to Hebe. Interestingly, the abundance of these Pop~III galaxies depends strongly on
the baryon-dark matter streaming velocity, $v_{\rm bc}$, because a higher
streaming velocity allows halos to remain chemically pristine for longer.
None of these late-forming Pop~III sources occur for $v_{\rm bc}=0$, while many form at high $v_{\rm bc}$. Averaging over the expected distribution of streaming
velocities, we predict that of order four GN-z11-like galaxies would need to
be surveyed at $z\sim11$ to expect to find one such source. We also note
these environments may provide a natural formation site for heavy black
hole seeds.}

\begin{document}
\maketitle
\flushbottom

\section{Introduction}
The search for Pop~III stars, defined by their extremely low metallicity, is currently an exciting frontier of astrophysics. Pop~III stars are first expected to form in low-mass minihalos with virial masses of ${\sim}10^{5-6}~M_\odot$ at $z\gtrsim30$ \citep[e.g.,][]{1999ApJ...527L...5B, 2002Sci...295...93A, 2003ApJ...592..645Y, 2009Sci...325..601T, 2010MNRAS.403...45S, 2015ComAC...2....3G}. Unfortunately, the very first Pop~III stars will be exceedingly difficult to detect due to their large cosmological luminosity distances and small predicted cluster masses \citep[e.g.,][]{2020MNRAS.492.4386S}. However, theoretical models have generically found that at least some Pop~III formation persists to $z{\sim}6$ or beyond (see Figure 2 in \cite{2023arXiv230312500K}). Pop~III stars at later times will be easier to detect, both because they are closer and because feedback from Lyman-Werner (LW) \citep{1997ApJ...476..458H, 2001ApJ...548..509M, 2008ApJ...673...14O,2014MNRAS.445..107V} and hydrogen ionizing radiation \citep{1998MNRAS.296...44G, 1996ApJ...465..608T, 2004MNRAS.348..753S, 2004ApJ...601..666D, 2006MNRAS.371..401H, 2014MNRAS.444..503N} push Pop~III star formation to higher halo masses (${\sim}10^{7-8}~M_\odot$), leading to a larger supply of gas to form stars.

Recently, the \emph{James Webb Space Telescope} (\emph{JWST}) has identified several such sources. This includes LAP1-B, which is consistent with a Pop~III galaxy with ${\sim}1000~M_\odot$ of massive Pop~III stars strongly magnified by gravitational lensing from a foreground galaxy cluster \citep{2023arXiv230514413V, 2025arXiv250611846N, 2025arXiv250803842V}. There is also Hebe, which shows a strong HeII 1640~\AA ~line and no metal lines, consistent with $2\times 10^4~M_\odot - 6\times10^5~M_\odot$ of Pop~III stars \cite{2024A&A...687A..67M, 2026arXiv260320362M,2026ApJ..1003L..14R}. Interestingly, Hebe is found at a 3 kpc projected distance from the massive galaxy GN-z11, which is at a redshift of $z=10.6$ and has a total stellar mass of ${\sim}10^9~M_\odot$ \citep{2023ApJ...952...74T}. 

The close proximity of Hebe to GN-z11 points to an interesting theoretical question: do we expect Pop~III galaxies to form in close proximity to massive high-redshift galaxies, which themselves form in highly biased regions of the high-redshift Universe? There are several competing effects. On one hand, highly overdense regions are expected to have their star formation histories pushed to higher redshifts, driving early enrichment which may cut off Pop~III star formation at relatively early times. On the other hand, near bright galaxies the LW background is very high, which could increase the star formation efficiency such that sources like Hebe could form \cite{2026ApJ..1006...89J, 2026ApJ..1006...27J}. 

In this paper we address this question using a semi-analytic model of
Pop~III and metal-enriched star formation applied to a zoom-in N-body
simulation of a highly overdense region resembling the environment of
GN-z11. We adopt a semi-analytic model because it is not yet possible to
resolve mini-halo formation in cosmological hydrodynamic simulations of 
such large and overdense regions, while including all the relevant physical
process required to accurately model them. 
Our simulation resolves minihalos and follows the earliest Pop~III
stars that form within this GN-z11-like environment, which is necessary to
accurately track metal enrichment and determine the correct abundance
of Pop~III stars at later times. Previous simulation work has suggested that
Pop~III stars may form in the outskirts of more evolved galaxies, but the
resolution of these simulations was insufficient to follow the early stages
of metal enrichment \citep{2023MNRAS.522.3809V}. 

We find that overdense regions like this
one do indeed give rise to a distinct population of late-forming, unusually
massive Pop~III sources, analogous to Hebe, and that their formation is
governed by the interplay between radiative feedback and the baryon-dark matter streaming velocity, $v_{\rm bc}$ \cite{2010PhRvD..82h3520T}. We
use our model to quantify the abundance of these sources as a function of
model parameters, and find that searches in the vicinity of massive,
high-redshift galaxies like GN-z11 offer a promising strategy for detecting
newly formed Pop~III stars.

This paper is organized as follows. In Section~2 we 
describe our methods, including our zoom-in N-body simulation inspired by GN-z11 and the details of our semi-analytic model. In Section~3 we present our 
results. In Section~4 we discuss these findings and present our 
conclusions. Throughout this work, we assume a $\Lambda {\rm CDM}$ cosmology with parameters consistent with \cite{2020A&A...641A...6P}: $\Omega_{\rm m} = 0.32$, $\Omega_{\Lambda} = 0.68$, $\Omega_{\rm b} = 0.049$, $h=0.67$, $\sigma_8=0.81$, and $n_{\rm s} = 0.96$. 

\section{Methods}
\subsection{N-body Simulation and Merger Trees}
We begin by describing the zoom-in N-body simulation which acts as the backbone of our semi-analytic
model. All of our simulations were performed with GADGET-4
\citep{2021MNRAS.506.2871S}, with initial conditions generated using MUSIC2
\citep{2011MNRAS.415.2101H, 2020ascl.soft08024H}. We first performed a dark-matter-only cosmological simulation of a 100 Mpc width box with $256^3$ particles, corresponding to a particle mass of $2.38 \times 10^9 \ M_\odot$. We identified the most massive halo at
$z = 11$ and resimulated its Lagrangian region  at high resolution with a
particle mass of $9060\ M_\odot$. Note that our high-resolution region was designed to extend beyond the halo itself so that we can study its environment (initial comoving width of 3.73 Mpc). 

This halo reaches $M_{\rm vir} \approx 8 \times 10^{10} \ M_\odot$ at
$z = 11$, which is similar to the halo mass inferred for GN-z11.
GN-z11 has an estimated stellar mass of $\approx 10^{9} \ M_\odot$ \citep{2023ApJ...952...74T}; matching this with our halo requires a star
formation efficiency of $M_\ast / [(\Omega_b/\Omega_m) M_{\rm vir}] \approx
0.08$, similar to the fiducial value of $f_{\rm II}=0.05$ adopted in our fiducial semi-analytic model (as described in the following subsection). Thus, we regard this halo as a reasonable GN-z11 analog, but emphasize that we only simulate a single such
region and defer a large statistical sample to future work.

\begin{figure}
\centering 
\includegraphics[width=15 cm]{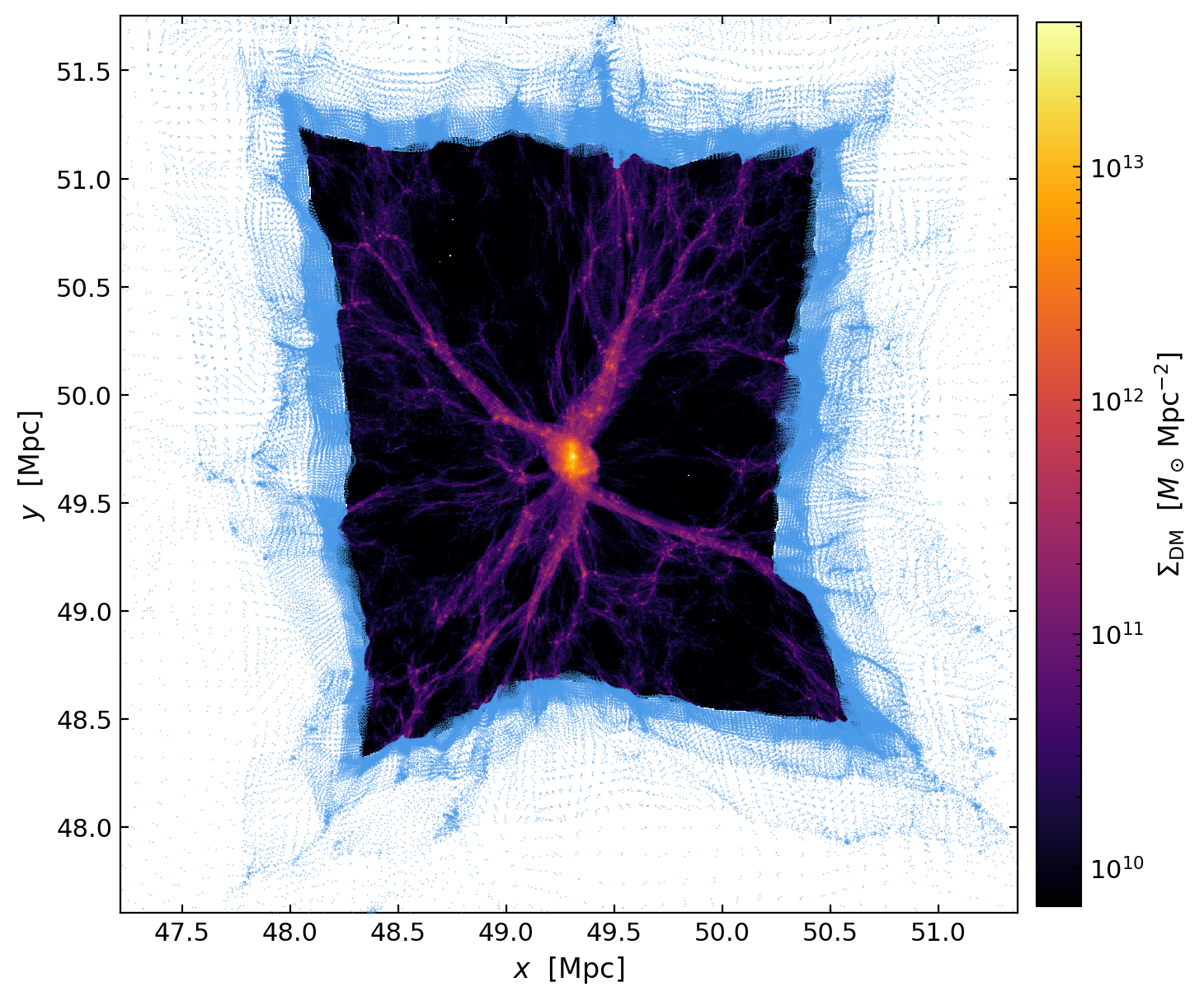}
\caption{\label{fig:zoom} Projected dark matter surface density in a slab of
thickness $0.21$ Mpc through the refined region of the zoom simulation at
$z = 11$. The projection is built from the high-resolution particles only and the fainter blue points around the edges are the low-resolution boundary
particles. This zoom targets the largest halo at $z=11$ in the 100 Mpc across parent simulation. }
\end{figure}

Initial conditions for the zoom-in simulation were generated at $z = 200$.
Particles within the Lagrangian region of the target halo were replaced with a higher-resolution set, and the resulting refined region was set to a cube
$3.73~{\rm Mpc}$ across, embedded in the full parent volume sampled with
progressively coarser particles. We saved approximately 100 snapshots between
$z = 60$ and $z = 11$, spaced such that the interval between consecutive
outputs is one tenth of the free-fall time at the virial
overdensity. In Figure~\ref{fig:zoom} we show the projected dark matter
surface density in a thin slab through the refined region at the final
snapshot. The high-resolution particles trace a network
of filaments converging on the central halo, and most halos lie along these
filaments and at their intersection. The low-resolution boundary particles surround the refined region, but there is essentially no contamination from low-resolution particles in the vicinity of the resolved halos.

The high-resolution particle mass quoted above is sufficient to resolve halos
at the minimum halo mass hosting Pop~III star formation, $M_{\rm crit}$, for $v_{\rm bc} = 1\sigma$ and $J_{\rm LW} = 0$. Following
\cite{2021ApJ...917...40K}, $M_{\rm crit} \approx 5 \times 10^5 \ M_\odot$ for this streaming velocity and LW flux at
$z \approx 30$, corresponding to $\approx 55$ particles. This resolution is sufficient to determine halo mass and position accurately \citep{2011MNRAS.415.2293K}. We note that for the lowest streaming velocity we consider, $v_{\rm bc} = 0$, halos hosting Pop~III star formation lie closer to our resolution limit, potentially causing an artificial delay for star formation (which could reduce metal enrichment leading to additional late-forming Pop~III star formation).

Halo catalogs were computed with the SUBFIND-HBT algorithm, and merger trees
were constructed using GADGET-4's merger tree builder
\citep{2021MNRAS.506.2871S}. We adopt the default SUBFIND-HBT
halo masses rather than the top-hat overdensity criterion of \cite{1998ApJ...495...80B}
used in our earlier work. While the two definitions give similar results, the
top-hat mass fails to converge for a small fraction of halos, whereas the
SUBFIND-HBT bound mass is well defined throughout the simulation. We only consider distinct halos (i.e., not subhalos), consistent with our treatment in \cite{2020ApJ...897...95V}.

\subsection{Semi-analytic Model}
Here we review the key features of our semi-analytic model, however see \cite{2020ApJ...897...95V} for a complete description. Generally speaking, our model takes the merger history and three-dimensional spatial information of dark matter halos from cosmological N-body simulations as input and utilizes analytic prescriptions to determine where Pop~III and metal-enriched star
formation occur. In particular, star formation and feedback processes are characterized by a number of model parameters (see Table \ref{tab:params}). Pristine halos form Pop~III stars with
efficiency $f_{\rm III}$ once they first reach the minimum mass for star
formation, $M_{\rm crit}$. For simplicity we do not follow the masses of individual stars and only assign a total stellar mass. After a delay $t_{\rm delay}$, subsequent star
formation in that halo proceeds as metal-enriched star formation with
efficiency $f_{\rm II}$ (which is calibrated to match observations of the UV luminosity function at $z{\sim}6$). Below the atomic cooling threshold (the halo mass corresponding to a virial temperature of ${\sim}10^4~{\rm K}$ where gas can cool via atomic H transitions), $M_{\rm crit}$ is set by molecular hydrogen cooling and follows the fitting functions of \cite{2021ApJ...917...40K}, which give the critical mass as a function of the local LW intensity and the
baryon-dark matter streaming velocity $v_{\rm bc}$. In our fiducial model we adopt twice the critical mass
given by these fits, motivated by comparison to the hydrodynamical simulation
AEOS \citep{Brauer_2025, 2025arXiv251011629H}. In regions that have been previously ionized, the gas is photoheated to ${\sim}10^4~{\rm K}$ and star formation is suppressed in halos below $M_{\rm ion}$ \citep{1996ApJ...465..608T, 1998MNRAS.296...44G, 2004MNRAS.348..753S, 2004ApJ...601..666D, 2006MNRAS.371..401H, 2014MNRAS.444..503N}.

Additionally, we reduce the ionizing escape fractions, $f_{\rm esc,II}$ and $f_{\rm esc,III}$, compared to our previous work \citep{2020ApJ...897...95V}. Because our refined region is overdense (we measure $1+\delta \approx 2.3$ during the epoch when reionization completes, $z\approx15\text{--}17$), it contains more hydrogen atoms than a mean-density volume of the same size, and therefore requires a correspondingly larger photon budget to reach the same ionization state. We reduce $f_{\rm esc,II}$ and $f_{\rm esc,III}$ by approximately this factor, to $0.04$ and $0.2$ respectively, to account for this. 

\begin{table}
\centering
\caption{Fiducial parameters of our semi-analytic model.}
\label{tab:params}
\begin{tabular}{lll}
\hline
Parameter & Description & Fiducial Value \\
\hline
$v_{\rm bc}$          & Streaming velocity at recombination           & $1.0\,\sigma$ (30 km/s)  \\
$f_{\rm bub}$         & Metal bubble size/speed scaling                & $0.4$ \\
$f_{\rm esc,II}$      & Metal-enriched ionization escape fraction          & $0.04$ \\
$f_{\rm esc,III}$     & Pop~III ionization escape fraction        & $0.2$ \\
$f_{\rm II}$          & Metal-enriched star formation efficiency       & $0.05$ \\
$f_{\rm III}$         & Pop~III star formation efficiency       & $0.001$ \\
$\eta_{\rm II}$       & Metal-enriched ionizing/LW photons per baryon & $4000$ \\
$\eta_{\rm III}$      & Pop~III ionizing/LW photons per baryon & $65000$ \\
$t_{\rm delay}$       & Delay between Pop~III and metal-enriched stars          & $10^{7}\,{\rm yr}$ \\
$Z_{\rm crit}$        & Critical metallicity   & $3\times10^{-4}\,Z_\odot$ \\
$M_{\rm ion}$         & Ionization feedback mass         & $1.5\times10^{8} \left (\frac{1+z}{11} \right )^{-3/2}\,M_\odot$ \\
\hline
\end{tabular}
\end{table}

We use a grid-based method to rapidly compute the three
feedback processes relevant to Pop~III star formation: LW radiation,
photoheating from inhomogeneous reionization, and metal enrichment from
supernova winds. We lay a $256^3$ grid over a cubic region 4.4 Mpc on a side,
chosen to span a region beyond the minimum and maximum positions reached by any resolved halo
over the full course of the simulation. We note that this grid volume is
larger than the 3.73 Mpc N-body refined region
described in Section 2.1. Our method assumes periodic boundary conditions on
this grid. We find that only the ionized bubble surrounding the most massive
halo in our simulation grows large enough to be affected by this choice, and
this bubble ionizes the full region containing all of our resolved halos
before it begins to wrap around the box. Metal bubbles remain far smaller than the box scale throughout the simulation. We therefore conclude that our (strictly speaking unphysical) periodic boundary conditions do not impact our results.

The LW intensity incident on a given halo is $J_{\rm LW} = J_{\rm loc} +
J_{\rm bg}$, where $J_{\rm loc}$ is computed directly from resolved sources on
the grid described above, and $J_{\rm bg}$ is a background contribution from
sources beyond our simulation volume. Because our refined region is a highly
overdense environment, we cannot estimate $J_{\rm bg}$ self-consistently from
its own resolved sources: doing so would effectively assume that the entire
universe out to the $\sim100$ Mpc LW horizon is filled with regions of
comparable overdensity, substantially overestimating the true background. We
instead impose a background floor calibrated from the large-volume merger-tree
model of \citep{2024ApJ...962...62F}, which we verify against our control ensemble (described
below) at mean density. We note that the LW horizon smooths large-scale
fluctuations in this background, such that even the value appropriate to our
highly overdense region is expected to lie within a factor of $\sim2$ of the
cosmic mean \citep{2025JCAP...02..043F}.

We model metal enrichment of the intergalactic medium (IGM) by assuming that
supernova winds drive spherical metal bubbles around star-forming halos, which
begin expanding 4 Myr after the onset of star formation. The bubble velocity is
$v_{\rm bub} = f_{\rm bub}\,60\ {\rm km\,s^{-1}}$ until the bubble reaches a
comoving radius $R_{\rm bub} = f_{\rm bub}\,150\,h^{-1}\ {\rm kpc}$; the free
parameter $f_{\rm bub}$ therefore scales both the expansion speed and the
maximum size of each metal bubble. We assume $10\ M_\odot$ of metals are
produced per $40\ M_\odot$ of Pop~III stars formed, and $1\ M_\odot$ per
$100\ M_\odot$ of metal-enriched stars formed.  Following \cite{2026JCAP...02..077V}, we also assume that half of metals produced by supernovae escape the halos and enter IGM bubbles.
Metals are distributed uniformly
within each bubble and summed where bubbles overlap, and this field is
computed on the same $256^3$ grid described above. A halo with mass above
$M_{\rm min,met} = 5\times 10^5~M_\odot$ that lies in a region enriched above $Z_{\rm crit}$ 
is then triggered to undergo metal-enriched star
formation. Compared to \cite{2020ApJ...897...95V}, we adopt a lower $f_{\rm bub}$ (reducing from 1.0 to 0.4), which we find agrees more closely with the metal bubbles in the AEOS simulation. 

To assess the significance of the environmental overdensity of our target
region, we construct a comparison ensemble of ten independent 3 Mpc boxes at
mean density (taken from \citep{2020ApJ...897...95V}). Because these control
boxes are at the mean density of the Universe, we compute
the LW background self-consistently from the sources within each box, rather
than imposing the externally calibrated background used for our zoom-in
simulation. We likewise use the fiducial escape fractions from \cite{2020ApJ...897...95V},
$f_{\rm esc,II}=0.1$ and $f_{\rm esc,III}=0.5$, in the control boxes, since the correction described above does not apply. Note that in all of our runs (both the zoom and the control regions) we assume a constant streaming velocity across the simulation box. This is a good approximation since the streaming velocity does not vary across regions of ${\sim}3$ Mpc.

\section{Results}
\begin{figure}
\centering 
\includegraphics[width=15.5 cm]{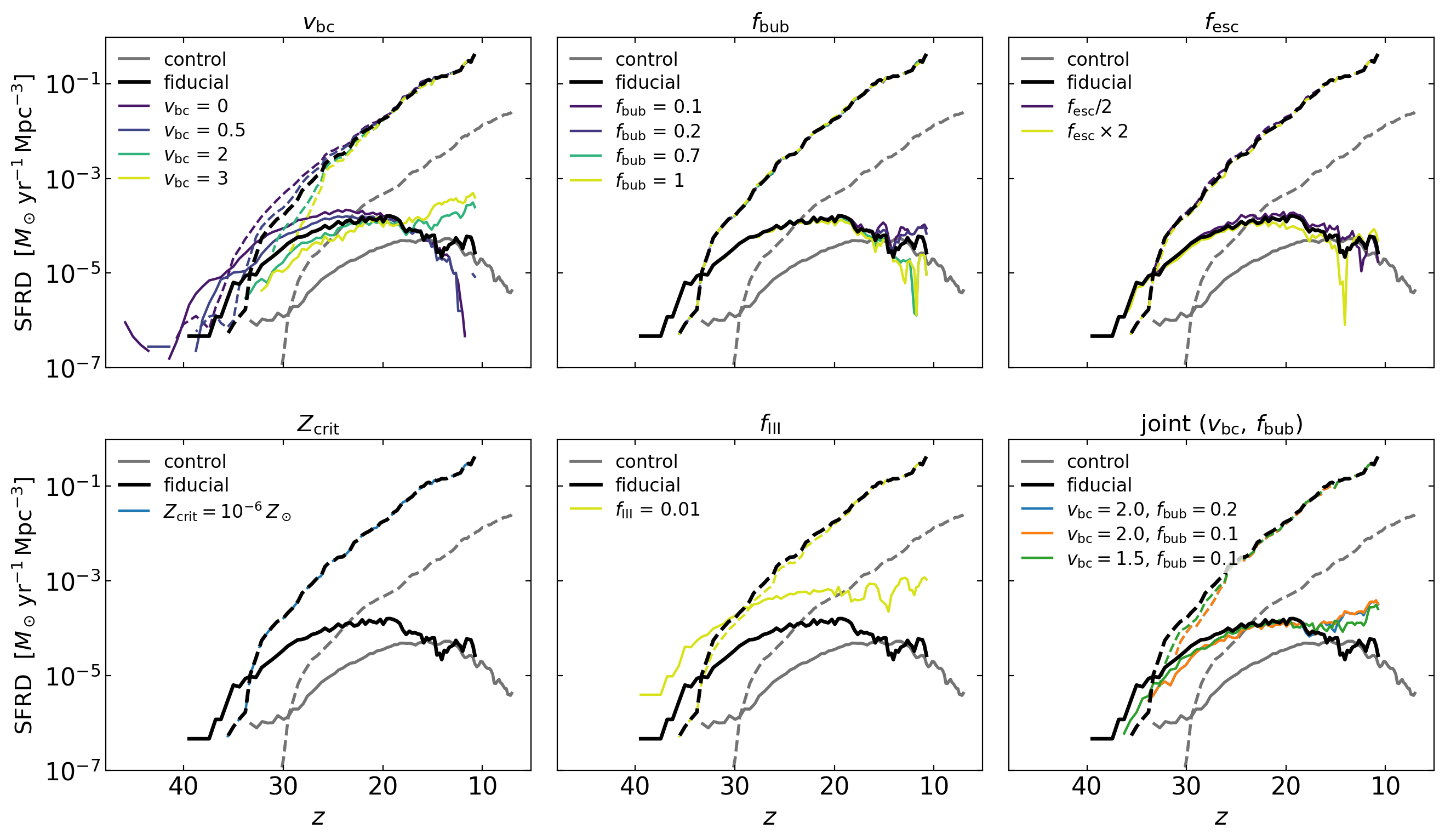}
\caption{\label{fig:sfrd} Pop~II (dashed) and Pop~III (solid) SFRD as a
function of redshift in the zoom region. We show the fiducial model as well as cases where one parameter has been individually varied while keeping the others at their fiducial values. Curves are smoothed with a trailing three-snapshot
running average.  The grey curves show the mean over ten $\delta = 0$ control
boxes to give a comparison with typical patches of the Universe.}
\end{figure}

In Figure~\ref{fig:sfrd} we show the Pop~III and metal-enriched star
formation rate density (SFRD) in our overdense zoom-in region compared to our
mean-density control ensemble. The overdense region is characterized by both
earlier and overall higher star formation compared to the control. We explore
our model's parameter space by varying a number of parameters around our
fiducial model. We find that most of these do not significantly change the
metal-enriched star formation history, with the exception of $v_{\rm bc}$,
which delays metal-enriched star formation to lower redshift. Regarding
Pop~III star formation, we find that it generally peaks near
$z\sim25$ and then gradually turns off. The $f_{\rm esc}$ and $Z_{\rm crit}$ parameters do
not strongly impact this behavior. The most interesting effect is that
increasing $v_{\rm bc}$ delays Pop~III star formation to lower
redshift, allowing Pop~III sources to be present near
GN-z11-like galaxies, as we discuss in more detail below. On the other hand, for
$v_{\rm bc}=0$, Pop~III star formation is almost entirely shut off by
$z\sim15$. Varying $f_{\rm bub}$ has a qualitatively similar effect, though
much more modest. We note that the volume used to compute the SFRD in our
zoom-in region is the comoving volume of the high-resolution region at high
redshift, $(3.73~{\rm Mpc})^3$, before it becomes somewhat compressed due to gravitational
collapse. 

\begin{figure*}
\centering
\includegraphics[width=\textwidth]{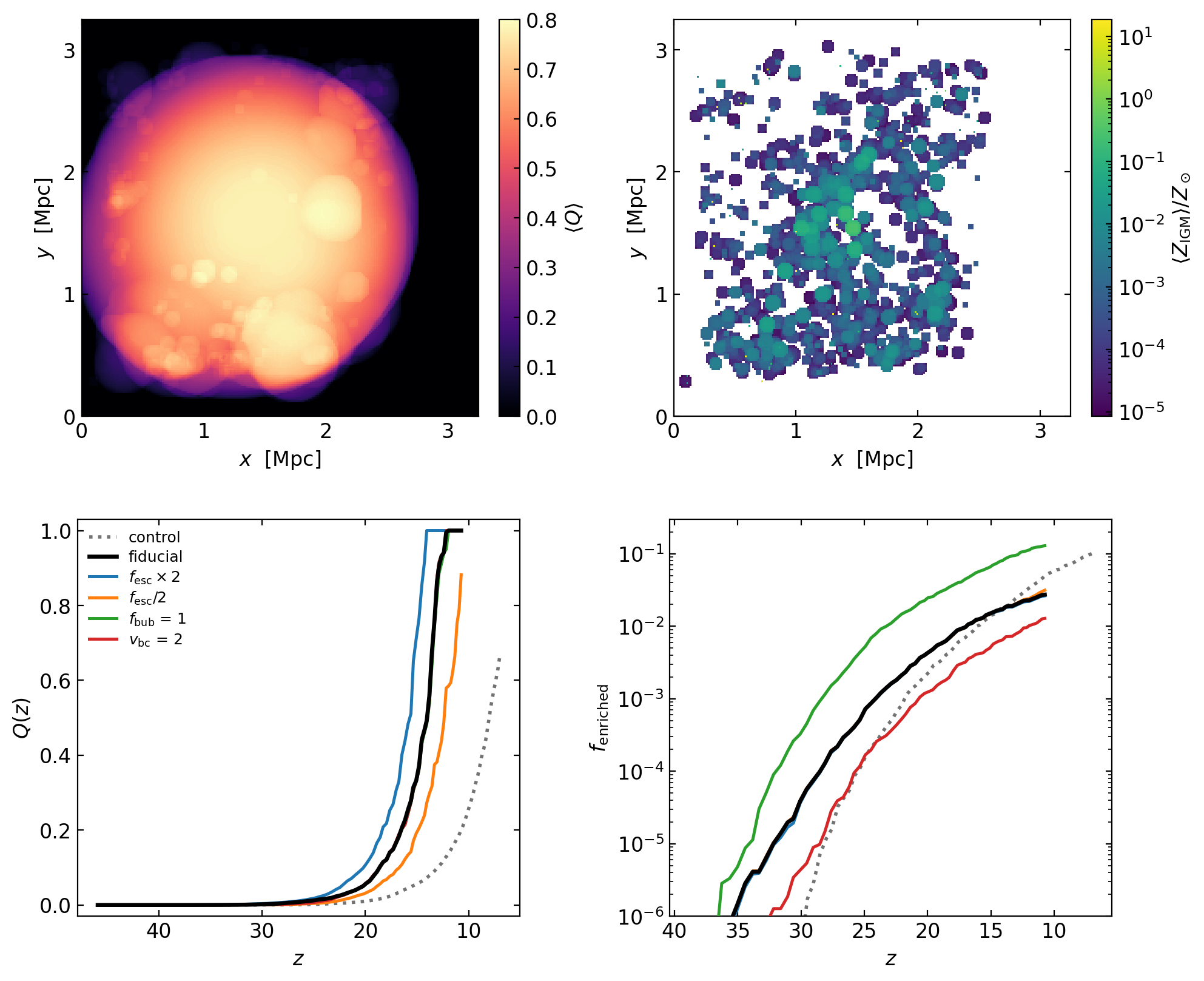}
\caption{\emph{Top panels:} projection of ionization (left) and metallicity (right) in our semi-analytic zoom-in simulation at $z=14.5$. \emph{Bottom panels:} ionized fraction (left) and volume fraction above $Z_{\rm crit}$ (right) in the refined region as a function of redshift.
\label{fig:bubbles}}
\end{figure*}

In Figure~\ref{fig:bubbles} we show projections of the ionization fraction
and IGM metallicity in our overdense region at $z{\sim}15$, along with the
evolution of the ionized and metal-enriched fractions of the zoom
region. Reionization within the zoom region is essentially complete
by this redshift, which subsequently shuts off star formation in halos below
$M_{\rm ion}$. In our fiducial model, only a few percent of the box is
enriched above $Z_{\rm crit}$ by $z{\sim}11$.

\begin{figure}
\centering 
\includegraphics[width=15 cm]{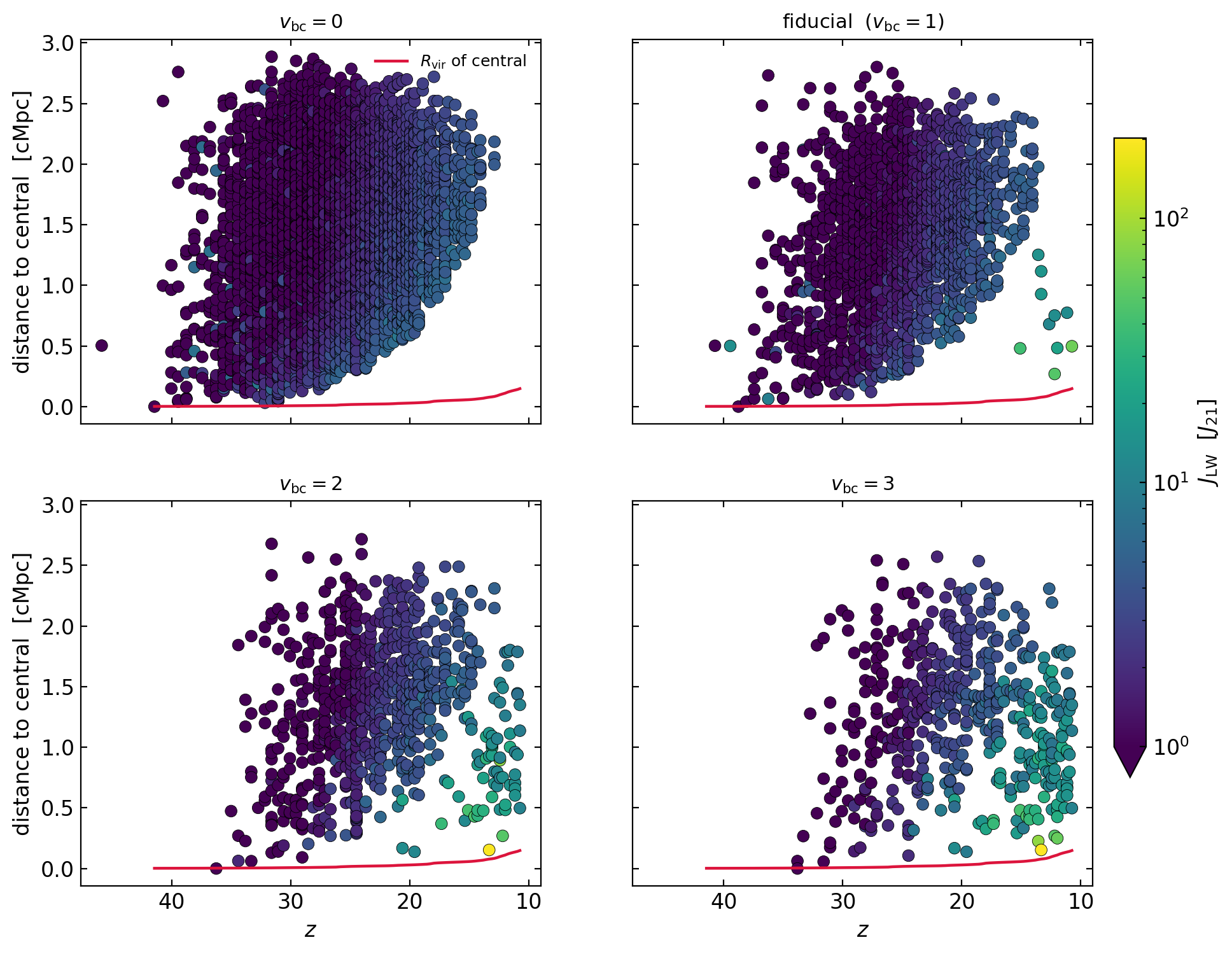}
\caption{\label{fig:dist} The comoving distance of newly formed Pop~III
sources measured from the main progenitor of the central GN-z11 analog as
a function of the redshift at which each Pop~III source forms. Points are
colored by the local LW intensity, $J_{21} \equiv J_{\rm LW}/(10^{-21}\ {\rm
erg\,s^{-1}\,cm^{-2}\,Hz^{-1}\,sr^{-1}})$. Panels show model
parameterizations with varying $v_{\rm bc}$ (all other parameters are held
at their fiducial values). The red curve traces the virial radius of the
central halo's main progenitor.}
\end{figure}

\begin{figure}
\centering 
\includegraphics[width=15 cm]{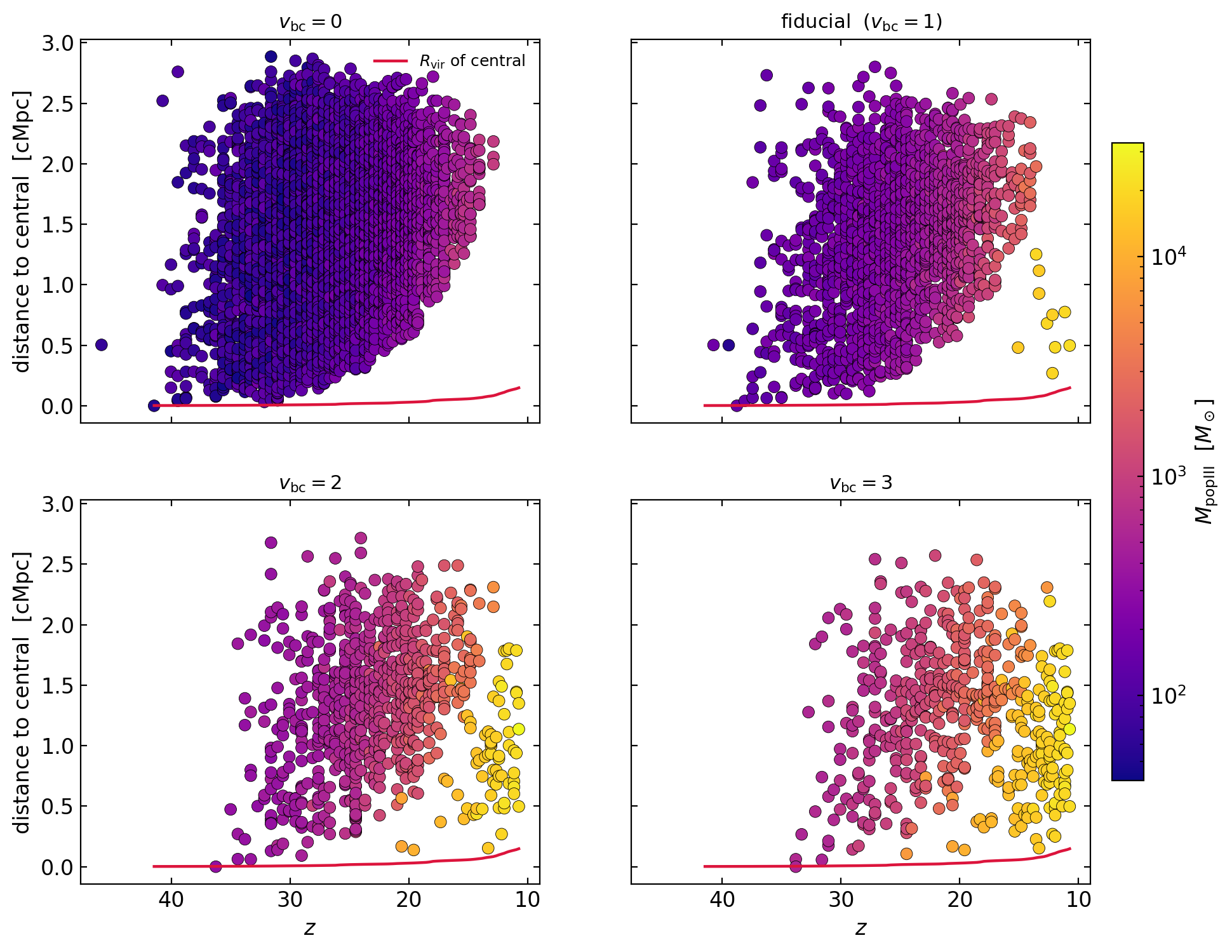}
\caption{\label{fig:dist_popIII} Same as Figure~\ref{fig:dist}, but indicating Pop~III mass with the color bar.  }
\end{figure}

Next, we examine the properties of Pop~III sources in our overdense region.
In Figures~\ref{fig:dist} and \ref{fig:dist_popIII} we show the distance from
the central GN-z11-like halo to newly formed Pop~III sources as a function
of redshift, colored by the local LW intensity $J_{21}$ and by the resulting
Pop~III stellar mass, respectively. Two populations are apparent. Early,
low-mass Pop~III sources are spread throughout the box, but as time
proceeds, the ionized region surrounding the central halo grows, allowing
only sources at progressively larger distances to exceed the critical mass.
A second, distinct population of late-forming Pop~III sources appears at
lower redshift. These are halos that are ionized but remain chemically
pristine and continue to accumulate mass until they exceed $M_{\rm ion}$,
producing a massive Pop~III starburst \cite{2017MNRAS.469.1456V, 2019ApJ...882..178K}.

These late-forming sources are plausible analogs to Hebe, the candidate
Pop~III source associated with GN-z11. They typically form in regions with
$J_{21}\sim10$--$100$, significantly higher than the $J_{21}\sim1$ typical of
our control ensemble. Such elevated $J_{21}$ has been shown to result in
either more massive Pop~III clusters or supermassive Pop~III stars
\citep{2026ApJ..1006...89J, 2026arXiv260813656H}. This suggests that overdense environments like the one studied
here give rise to a distinct class of Pop~III galaxies more massive
than is typical in mean density regions of the Universe. Their formation
requires halos to remain pristine despite the surrounding overdensity, which
we find results from a high streaming velocity, $v_{\rm bc}$, as we discuss
in detail below. For $v_{\rm bc}=0$ we see essentially none of these massive
late-forming Pop~III sources, but $\sim10$ form for $v_{\rm bc}=1$ and many more
at higher $v_{\rm bc}$. We note that some of these sources lie very close to
the virial radius of the central galaxy, but since our semi-analytic model
does not currently include subhalos, we cannot directly probe sources at
very small separations, such as the $3$ kpc projected distance between Hebe
and GN-z11.

\begin{figure}
\centering 
\includegraphics[width=15 cm]{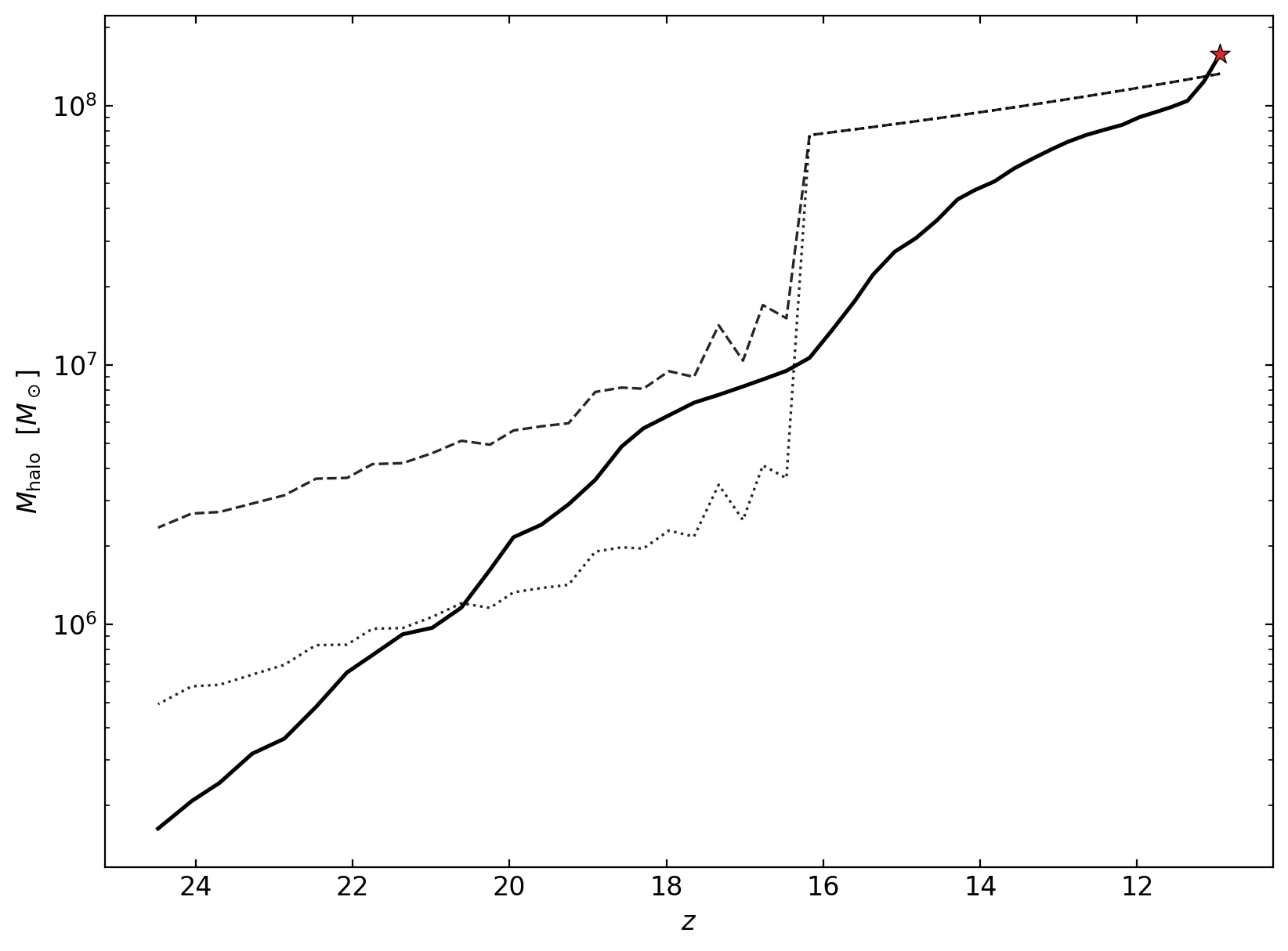}
\caption{\label{fig:halo_tracks} The halo mass of the main-branch progenitor of a halo hosting a
late-forming Pop~III galaxy in our $v_{\rm bc}=1.5\,\sigma$ run (solid
curve), together with the critical halo mass for star formation, based on
this halo's own $J_{\rm LW}$ history and time of local reionization
(dashed curve). The red star indicates the time of Pop~III star formation. This halo forms Pop~III stars late because $M_{\rm crit}$ has been
raised by the combined effect of LW feedback and reionization (the latter produces the late jump in the critical mass). We also show what $M_{\rm crit}$
would be for this same halo if the streaming velocity were zero (dotted
curve). In this case, the halo would first form stars at $z\sim20$,
illustrating how streaming velocities can delay Pop~III star formation
to lower redshift, where the halo is more massive.}
\end{figure}

In Figure~\ref{fig:halo_tracks} we illustrate the mechanism responsible for
these late-forming, massive Pop~III sources by tracking the mass evolution
of the main branch progenitor of a representative late-forming Pop~III halo in
our $v_{\rm bc}=1.5\,\sigma$ run, alongside the critical mass, $M_{\rm crit}$,
that the halo must exceed to form Pop~III stars. $M_{\rm crit}$ is initially
set by the combination of $J_{\rm LW}$ and $v_{\rm bc}$ following
\cite{2021ApJ...917...40K}, but rises sharply once the halo's local environment is
reionized, at which point $M_{\rm crit}$ instead follows $M_{\rm ion}$. This
combination allows the halo to remain pristine to much lower redshift than it
otherwise would, so that once it finally exceeds $M_{\rm crit}$, it does so
at high mass, producing a correspondingly massive burst of Pop~III star
formation. As discussed above, this delay depends sensitively on
$v_{\rm bc}$. For $v_{\rm bc}=0$, $M_{\rm crit}$ for this halo is already low prior to
reionization, so its main progenitor would cross this threshold much earlier and the massive late-forming Pop~III galaxy would not emerge. This behavior is shared by other late-forming Pop~III galaxies in our zoom simulation.

\begin{figure}
\centering 
\includegraphics[width=15 cm]{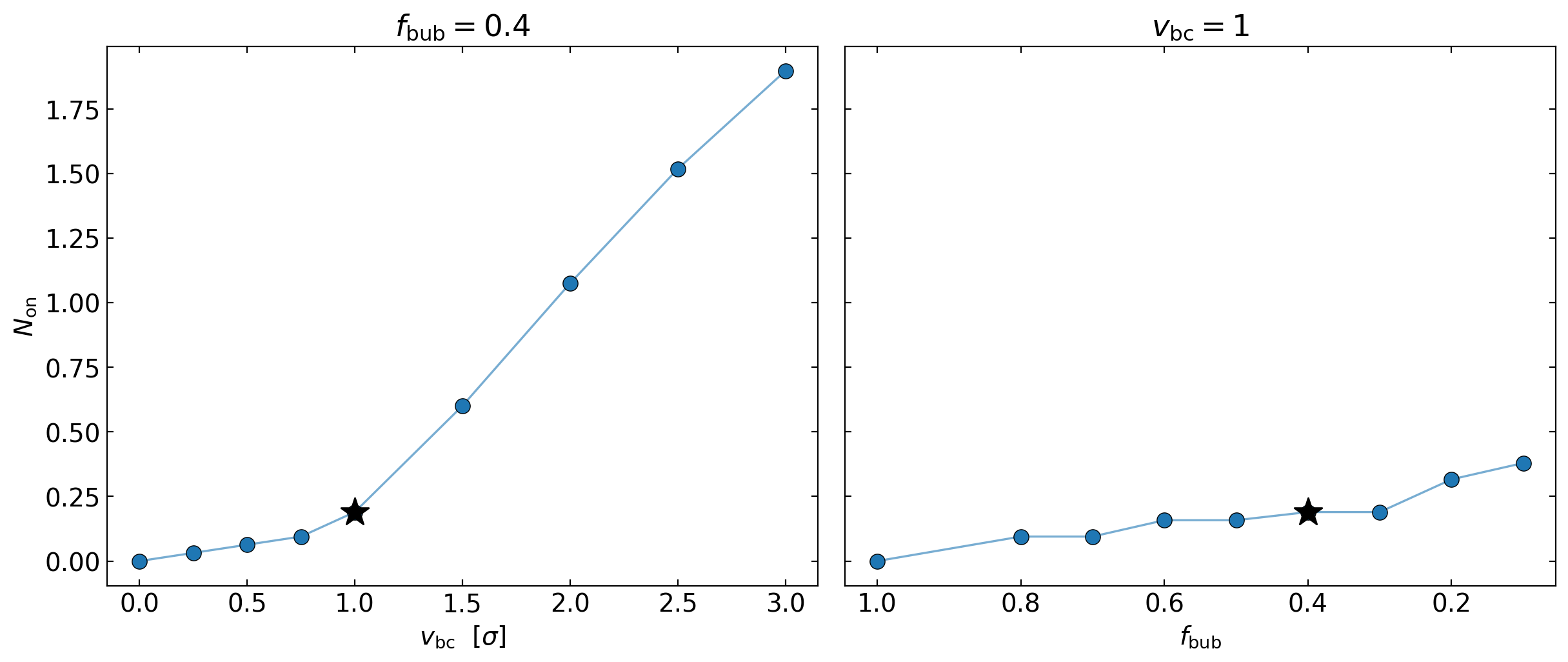}
\caption{\label{fig:N_on} Expected number of simultaneously luminous Pop~III sources, $N_{\rm on}$, in the refined region at the end of the simulation ($z=11-12$), as a function of streaming velocity (left, at fiducial $f_{\rm bub}$) and of the metal bubble velocity and maximum size (right, at fiducial $v_{\rm bc}$). The fiducial value, $N_{\rm on}$ is represented in both panels with a star.}
\end{figure}

Having established the mechanism responsible for these late-forming, massive
Pop~III sources, we next quantify their predicted abundance as a function of
$v_{\rm bc}$ and $f_{\rm bub}$. In Figure~\ref{fig:N_on} we plot the number of
Pop~III sources expected to be visible at $z=11$--$12$, computed as
$N_{\rm on} = N_{\rm formed}\,t_{\rm life}/t_{\rm interval}$, where
$N_{\rm formed}$ is the number of Pop~III sources that form within this
redshift interval, $t_{\rm interval}\approx50$ Myr is the cosmic time spanned
by $z=11$--$12$, and we adopt a Pop~III lifetime of $t_{\rm life}=3$ Myr. We
find that $N_{\rm on}$ depends strongly on $v_{\rm bc}$, increasing from
$\sim0$ to $\sim2$ as $v_{\rm bc}$ increases from $0$ to $3\,\sigma$. The
dependence on $f_{\rm bub}$ is comparatively weak, with $N_{\rm on}$
increasing from $\sim0$ to only $\sim0.4$ as $f_{\rm bub}$ decreases from
$1.0$ to $0.1$. At our fiducial parameters, $N_{\rm on}=0.19$. Averaging over
the Maxwell-Boltzmann distribution of streaming velocities, we
find a mean value of $\langle N_{\rm on}\rangle = 0.24$. We note that this
estimate is conservative, in the sense that our model does not currently
allow subhalos of the central galaxy to form stars; we defer a treatment of
this additional source of Pop~III star formation to future work.

\section{Discussion and Conclusions}
In this paper, we utilized a semi-analytic model to analyze a ${\sim} 4~{\rm Mpc}$ overdense region surrounding the most massive halo in a 100 Mpc cosmological box. This halo reaches a mass of $8\times10^{10}~M_\odot$ at $z=11$ making it an analog of GN-z11. Our semi-analytic model follows the formation of Pop~III and metal-enriched stars including important physics such as LW feedback, the baryon-dark matter streaming velocity, as well as ionization and metal enrichment of the IGM. In our model the buildup of the massive halo was followed by an N-body zoom-in simulation which resolves the minihalos where the first stars form.

We find that Pop~III and metal-enriched star formation both occur much
earlier and with a much higher SFRD in our overdense region compared to a
typical, mean-density environment. In our fiducial model, the majority of
Pop~III star formation peaks near $z\sim25$ and gradually diminishes by
$z\sim15$, as halos are enriched with metals. However, we find a distinct
population of late-forming Pop~III galaxies: halos that are reionized before
they reach sufficient mass to form Pop~III stars, remain pristine and
continue accumulating mass until they exceed $M_{\rm ion}$, at which point
they undergo a massive burst of Pop~III star formation. These late-forming
sources preferentially form closer to the central galaxy, and consequently
experience a much higher LW intensity ($J_{21}\sim10$--$100$) than typical
Pop~III sources in a mean-density environment. Owing to their higher halo
masses and elevated $J_{21}$, these late-forming sources are likely to
produce unusually massive Pop~III galaxies. We therefore predict a
population of massive Pop~III galaxies in the vicinity of massive
high-redshift galaxies like GN-z11, analogous to Hebe.

The number of massive, late-forming Pop~III galaxies in our model depends
strongly on the streaming velocity, $v_{\rm bc}$, of the region, with higher
$v_{\rm bc}$ producing many more such sources. This occurs because a higher
$v_{\rm bc}$ raises the minimum halo mass for Pop~III star formation,
allowing halos to remain pristine up to, and past, the point at which they
are ionized by the central source. We predict that, near a galaxy like
GN-z11, $\langle N_{\rm on}\rangle=0.24$ of these sources will be visible as
newly formed Pop~III galaxies at any given time near $z\sim11$, meaning that
roughly four such galaxies would need to be surveyed to expect to find one.
We note that this estimate is conservative, in the sense that our model
does not allow for star formation within subhalos. Hebe is likely
to be hosted by a subhalo given its close proximity to GN-z11. 

The parameter $f_{\rm bub}$ also impacts the abundance of these sources, but this is a much more modest effect. Our fiducial maximum metal bubble radius corresponds to a physical size of ${\sim}8$ kpc at $z=11$, similar to the $\sim10$ kpc bubble sizes found around $\sim10^{11}\,M_\odot$ halos in the analytic wind model of \citep{2003ApJ...588...18F}, suggesting our fiducial bubble sizes are not unphysically small relative to the mass of our central GN-z11 analog. We additionally explore larger bubble sizes by increasing $f_{\rm bub}$ up to a value of 1.

Our findings also have interesting implications for supermassive black hole
(SMBH) formation. A proposed pathway to SMBH formation involves so-called
heavy seeds, black holes formed from the collapse of ${\sim}10^{4-5}~
M_\odot$ supermassive stars \citep{2020ARA&A..58...27I}. These massive black hole
seeds may help explain the emergence of abundant SMBHs already in place in
the early Universe. Recent simulations have shown that pristine halos
exposed to $J_{21}\gtrsim1$--$10$ may form supermassive stars rather than a normal Pop~III
cluster \citep{2026arXiv260813656H}. Our late-forming Pop~III galaxies meet this criterion,
suggesting that in regions of moderate to high $v_{\rm bc}$, massive
galaxies like GN-z11 are natural locations for heavy black hole seeds to
form.

Several caveats should be taken into account when interpreting these results. We
have simulated only a single realization of a GN-z11-like environment, and
so cannot characterize the scatter expected among such regions. Our model
does not currently include star formation within subhalos of the central
galaxy, making our predicted abundance of late-forming Pop~III sources
conservative and making it difficult to compare directly with Hebe. In
future work, we plan to extend our model to include star formation within
subhalos, simulate a larger statistical sample of overdense regions to
characterize environment-to-environment scatter, and explore a broader
range of environments, including different redshifts and degrees of
overdensity. Our results suggest that the vicinity of massive,
high-redshift galaxies like GN-z11 is a promising place to search for
Pop~III stars, and may represent a natural site for the formation
of the heavy black hole seeds required to explain the earliest supermassive
black holes.

\acknowledgments
EV acknowledges the support of NSF grant AST-2009309, NASA ATP grant 80NSSC22K0629, and STScI grant JWST-AR-05238. 
GLB acknowledges support from the NSF (AST-2307419), NASA TCAN award 80NSSC21K1053, and the Simons Foundation through the Learning the Universe Collaboration.
The authors acknowledge the use of Claude Sonnet, Claude Opus, and Claude
Fable (Anthropic) for assistance in editing portions of this manuscript and
in developing analysis code used in this work.

\bibliography{GNz_11}
\end{document}